\documentclass[aps, reprint]{revtex4-1}

\usepackage[utf8]{inputenc}
\usepackage{graphicx}
\usepackage{hyperref}
\usepackage{amsmath}

\begin{document}
	
	\title{Energy Filtering in Doping Modulated Nanoengineered Thermoelectric Materials: A Monte Carlo Simulation Approach}
	
	\author{Pankaj Priyadarshi}
	\email{pankaj.priyadarshi@warwick.ac.uk}
	\author{Vassilios Vargiamidis}
	\email{vasileios.vargiamidis@warwick.ac.uk}
	\author{Neophytos Neophytou}
	\email{n.neophytou@warwick.ac.uk}
	
	\affiliation{School of Engineering, University of Warwick, Coventry, CV4 7AL, United Kingdom \\}
	
	
	\begin{abstract}
		Using Monte Carlo electronic transport simulations, coupled self-consistently with the Poisson equation for electrostatics, we explore the thermoelectric power factor of nanoengineered materials. These materials consist of alternating highly doped and intrinsic regions on the scale of several nanometers. This structure enables the creation of potential wells and barriers, implementing a mechanism for filtering carrier energy. Our study demonstrates that by carefully designing the nanostructure, we can significantly enhance its thermoelectric power factor compared to the original pristine material. Importantly, these enhancements stem not only from the energy filtering effect that boosts the Seebeck coefficient but also from the utilization of high-energy carriers within the wells and intrinsic barrier regions to maintain relatively high electronic conductivity. These findings can offer guidance for the design and optimization of new-generation thermoelectric materials through improvements in the power factor.
	\end{abstract}
	\maketitle
	\section{Introduction}
	In the past several years, large efforts have been devoted to the field of thermoelectric (TE) materials, resulting in significant progress in their performance~\cite{Beretta2019,Zou2020,Shankar2023,Artini2023}. The~performance of TE materials is generally quantified by the figure of merit {$ZT$}, which is a measure of the ability of a material to convert heat into electricity. It is determined by\linebreak $ZT = \sigma S^2 T / ( \kappa_e + \kappa_l )$, where $\sigma$ is the electronic conductivity, $S$ is the Seebeck coefficient, $T$ is the absolute temperature, and~$\kappa_e ( \kappa_l )$ is the electronic (lattice) thermal conductivity. The~product $\sigma S^2$ is known as the power factor of a TE material. Over~the past decade, the~figure of merit has more than doubled, surpassing values of $ZT > 2$ in several materials and across various temperature ranges \cite{Beretta2019,Zou2020,Shankar2023,Artini2023,Wu2014,Biswas2012,Zhao2014,Fu2016,Liu2012,RKim2015,Olvera2017,Rogl2014,Hao2019,Fiorentini2019,Finn2021,Wei2020,Wolf2019}.
		
	\indent The improvement in the figure of merit primarily arises from the significant reduction in lattice thermal conductivity observed in nanostructured materials, reaching values approaching the amorphous limit of $\kappa_l = 0.1-2$ W/mK and~even lower in certain cases~\cite{Beretta2019,Biswas2012,Nakamura2018,Taborda2016,Lee2017,Ohnishi2019,Kearney2018}. For~instance, amorphous silicon typically has a thermal conductivity between 0.1 and 2 W/m-K, with~common values around 0.7 to 1.5 W/mK for thin films~\cite{Amirreza2020}. Similarly, amorphous germanium has a thermal conductivity between 0.1 and 1 W/mK, depending on the conditions and sample preparation~\cite{Jeffrey2016}. Hierarchical nanostructuring, a~highly successful approach for reducing thermal conductivity~\cite{Biswas2012,Zhao2014}, involves the introduction of embedded atomic defects, nanoinclusions, and~grain boundaries, leading to extensive phonon scattering and consequently lower thermal conductivity. These structural distortions scatter phonons across various wavelengths, effectively reducing phonon transport across the entire spectrum~\cite{Gordillo2023}. Nonetheless, despite achieving ultra-low thermal conductivities, sometimes falling well below the amorphous limit, further enhancements to $ZT$ are expected to stem primarily from improvements in the power~factor. 
	
\begin{figure*}
		\includegraphics[height=3.5cm, width=15.5cm ]{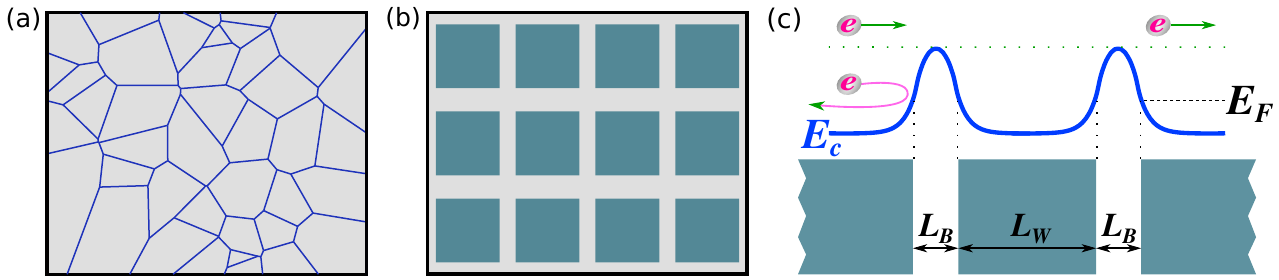}
		\caption{
			(\textbf{a}) A nanocrystalline material, having notably high thermoelectric power factors through high doping and the formation of potential barriers along grain boundaries. (\textbf{b}) A nanoengineered controlled version of (\textbf{a}), featuring regularly spaced highly doped square regions separated by undoped regions. {(\textbf{c})} A schematic 1D cross-section of the conduction band $E_c$ profile for the nanostructure depicted in (\textbf{b}), derived from the self-consistent solution of the Poisson equation. $L_W$ and $L_B$ denote the widths of the well and barrier, respectively, and~$E_F$ represents the Fermi level. Consequently, the~energy filtering features allow electrons at higher energies to propagate while hindering electrons at lower~energies, depicted with green and pink arrow respectively.}
		\label{fig1}
	\end{figure*}

	\indent The challenge in enhancing the power factor ($PF$) is attributed to the adverse interdependence between electronic conductivity and the Seebeck coefficient via the carrier density, which tends to keep the $PF$ low. However, a~promising direction to address this challenge lies in energy filtering techniques, achieved through the incorporation of energy barriers. These barriers effectively block low-energy carriers while permitting the flow of high-energy carriers~\cite{Bahk2014,Bahk2016,Zhao2014,NN2013,Vassilios2019,RKim2012,Gayner2020,Sakane2019,Ishibe2018,Liang2023}. Consequently, the~Seebeck coefficient (which is a measure of the energy of current flow) increases. Various methods for implementing energy filtering exist, including the use of superlattices~\cite{Bahk2014,Vashaee2004}, poly/nanocrystalline materials (as illustrated in Figure~\ref{fig1}a, where barriers are formed on the grain boundaries) \cite{NN_Nanotech_2013,Narducci_APL_2021}, and materials featuring dislocation loops~\cite{Bennett_2016}, among~others. Indeed, recent efforts focusing on employing energy filtering for the grain/grain-boundary systems across several materials have shown improved thermoelectric performance~\cite{Gayner2020,Sakane2019,Ishibe2018}. 
		
	\indent In this computational study, we investigate the TE performance of material systems in which the wells and energy filtering potential barriers are electrostatically formed by alternating heavily doped and intrinsic regions, as~depicted in Figure~\ref{fig1}b. This investigation is inspired by our previous work on such energy filtering systems, as~outlined in Ref.~\cite{NN_MatToday_2019}, utilizing simple series resistance models, as~well as experimental studies on nanocrystalline materials (Figure \ref{fig1}a). These studies revealed that ultra-high $PF$s can be achieved by forming energy barriers at grain boundaries, surpassing five times the optimal pristine material $PF$ \cite{NN_Nanotech_2013,Narducci_APL_2021}. 
	In Ref.~\cite{NN_MatToday_2019}, we presented the concept of achieving such PF improvements in detail. However, the~simple series resistance analytical model we used, by~its nature, includes many uncertainties, simplifications, and~assumptions with regard to the treatment of electronic and thermoelectric transport. The~purpose of our current study is to re-evaluate this design concept using a full-scale advanced Monte Carlo simulation formalism, which relaxes many of the approximations of the simplistic series resistance model and provides more confidence in the proposed design, a~capability that we did not have at the time of the original concept paper. Other than transport specifics, one important addition in this work is that we employ the self-consistent solution of the Poisson equation to obtain the accurate potential barriers and their shapes for a specific underlying doping~profile.
	
	\indent We employ Monte Carlo (MC) simulations to solve the semi-classical Boltzmann transport equation (BTE) in a 2D domain, based on an advanced method and code that we have developed specifically for efficient transport computation in nanostructures~\cite{Pankaj2023}. The~basic structure we consider is shown in Figure~\ref{fig1}b, with~blue domains representing highly doped regions and gray domains representing undoped regions. This approach is then coupled self-consistently with the 2D Poisson equation to capture the electrostatics of the domain and the shape of the energy band profile, $E_c$, that consists of electrostatic potential barriers and wells, as~depicted in Figure~\ref{fig1}c through a 1D schematic example.
	
	\indent Our computational framework accounts for phonon scattering, as~well as phonon plus ionized impurity scattering (IIS), with~the latter being particularly relevant in potential well regions. We then present a novel design for nanostructured materials capable of significant $PF$ enhancements. This design relies on several key concepts or ``ingredients'', whose contributions are comprehensively described in Refs.~\cite{NN_Nanotech_2013,Narducci_APL_2021}, and~involves the following: (i) energy filtering in the presence of potential barriers, (ii) reducing the well length to effectively transfer the Seebeck effect of the barrier into most of the well region (i.e., allowing carriers to propagate at high energies), (iii) high doping to utilize carriers with high velocities and position the Fermi level near the top of the potential barrier, and~(iv) an undoped barrier region to mitigate the reduction in electrical conductivity introduced by the barriers (referred to as the ``clean filtering'' approach~\cite{NN_MatToday_2019}). 
	
	\indent The paper is structured as follows: In Section \ref{sec2}, we introduce our approach and simulation method along with the key features of the design parameters. Section \ref{sec3} describes the $PF$ performance by investigating the improvements as a consequence of the design. We summarize our findings and conclude in Section~\ref{sec4}.
	
	\section{Method and~Approach}\label{sec2}
	
	In our approach, we introduce a well--barrier structure within a defined domain by selectively doping the domain in a periodic manner, as~depicted by the dark-colored square regions in Figure~\ref{fig1}b. This doping variation induces an electrostatic potential energy variation, which is computed using the Poisson equation,
	
	\begin{equation}
		\nabla^2 \phi = -\frac{\rho}{\epsilon_0 \epsilon_r},
		\label{eq1}
	\end{equation}
	where $\phi$ is the potential, $\rho$ is the charge density, and~$\epsilon_0$ ($\epsilon_r$) is the vacuum (relative) permittivity of the medium. The~charge density in the doped semiconductor is obtained~by
	\begin{equation}
		\rho = 	\int g(E) f (E ) dE,
		\label{eq2}
	\end{equation}
	where $g(E)$ is the density of states and $f(E)$ is the Fermi--Dirac distribution function. We then solve the Poisson equation self-consistently across the entire 2D domain (with dimensions of $110~\mathrm{nm} \times 90~\mathrm{nm}$), subject to Neumann boundary conditions at the boundaries. This computational process yields the charge density distribution and determines the profile of the conduction band energy $E_c$. Subsequently, the~$E_c$ profile serves as an input for the Monte Carlo ray-tracing simulation to extract the electron flux. The~term ``ray tracing'' is well-known in fields like particle physics and computer graphics for Monte Carlo simulations. But,~similarly, it has been adopted in the electronic transport community, primarily in the past, to trace the pathways for electrons in transistor devices in the presence of potential variations, boundaries, and~scattering events~\cite{Fischetti1988, Jacoboni1983}. Here, we have adopted this approach to investigate how charge carriers move through the nanostructured domains in real space and~use the term ``ray-tracing'' as a reference to keeping track of the electronic trajectories and the time electrons spend in the~domain.
	
\begin{figure}
		\includegraphics[height=8.4cm, width=3.8cm ]{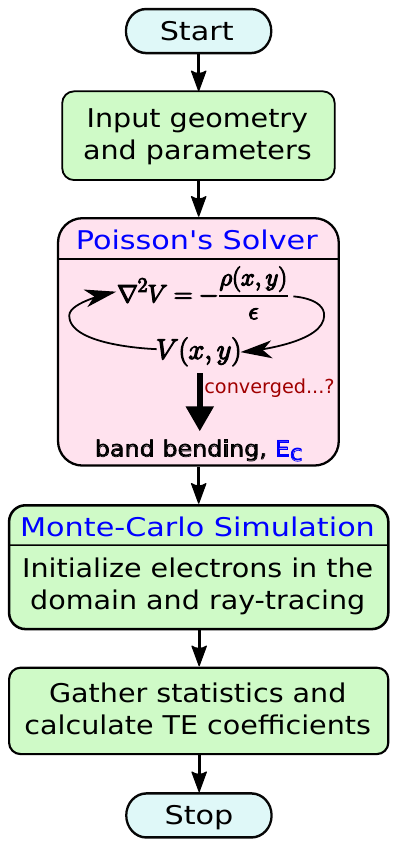}
		\caption{Flowchart that describes the calculation method for electron transport in the domain. It utilizes the self-consistent solution of the Poisson equation and a Monte Carlo ray-tracing algorithm, as~elaborated in Ref.~\cite{Pankaj2023}.}
		\label{fig2}
	\end{figure}

	\indent Our ray-tracing method, recently developed and detailed in Ref.~\cite{Pankaj2023}, is specifically tailored for intricate nanostructured materials and optimized for complex thermoelectric material geometries, ensuring computational efficiency. While comprehensive methodological details are presented in Ref.~\cite{Pankaj2023}, we provide a brief overview for completeness, as~depicted in the flowchart in Figure~\ref{fig2}.
		
	\indent Our method adopts a single-particle incident flux, where electrons are initialized sequentially at the left boundary of the domain and proceed either to the opposite side or undergo backscattering. Instead of randomly selecting free-flight times and implementing self-scattering as in previous Monte Carlo approaches~\cite{lundstrom}, we use a mean-free-path (mfp, $\lambda$) approach~\cite{Dhritiman2018}, where electrons propagate an mfp, followed by definite scattering after each mfp is completed. For~electron-phonon scattering, we assume an mfp of $\lambda_{ph}=30$ nm, and~for the bandstructure, we use a parabolic band effective mass $m^*=0.9m_0$. This choice leads to a low carrier density mobility of $n$-type Si of roughly 1500 cm$^2$/V.s (here, we consider only elastic phonon scattering for simplicity). We emphasize, though, that it did not attempt to simulate specific materials, as~this concept is material agnostic. We also calculate the IIS rate using the Brooks--Herring model combined with the strong screening scattering model in each discretization cell within the simulation domain, which aligns better with experimental mobility measurements, particularly for pristine Si~\cite{NN_Nanotech_2013,NN_MatToday_2019}. Using the band structure velocity, we determine the mfp ($\lambda_{IIS}$) due to IIS and~then employ Matthiessen's rule to compute the total mean free path. An~electron in a simulation domain cell advances one total mfp, $\lambda_{total}$, before~undergoing enforced scattering. Notably, because of the IIS process, $\lambda_{total}$ varies as a function of energy (although $\lambda_{ph}$ is generally energy-independent, at~least for acoustic phonon scattering).
	
\begin{figure*}
		\includegraphics[height=10.5cm, width=15cm ]{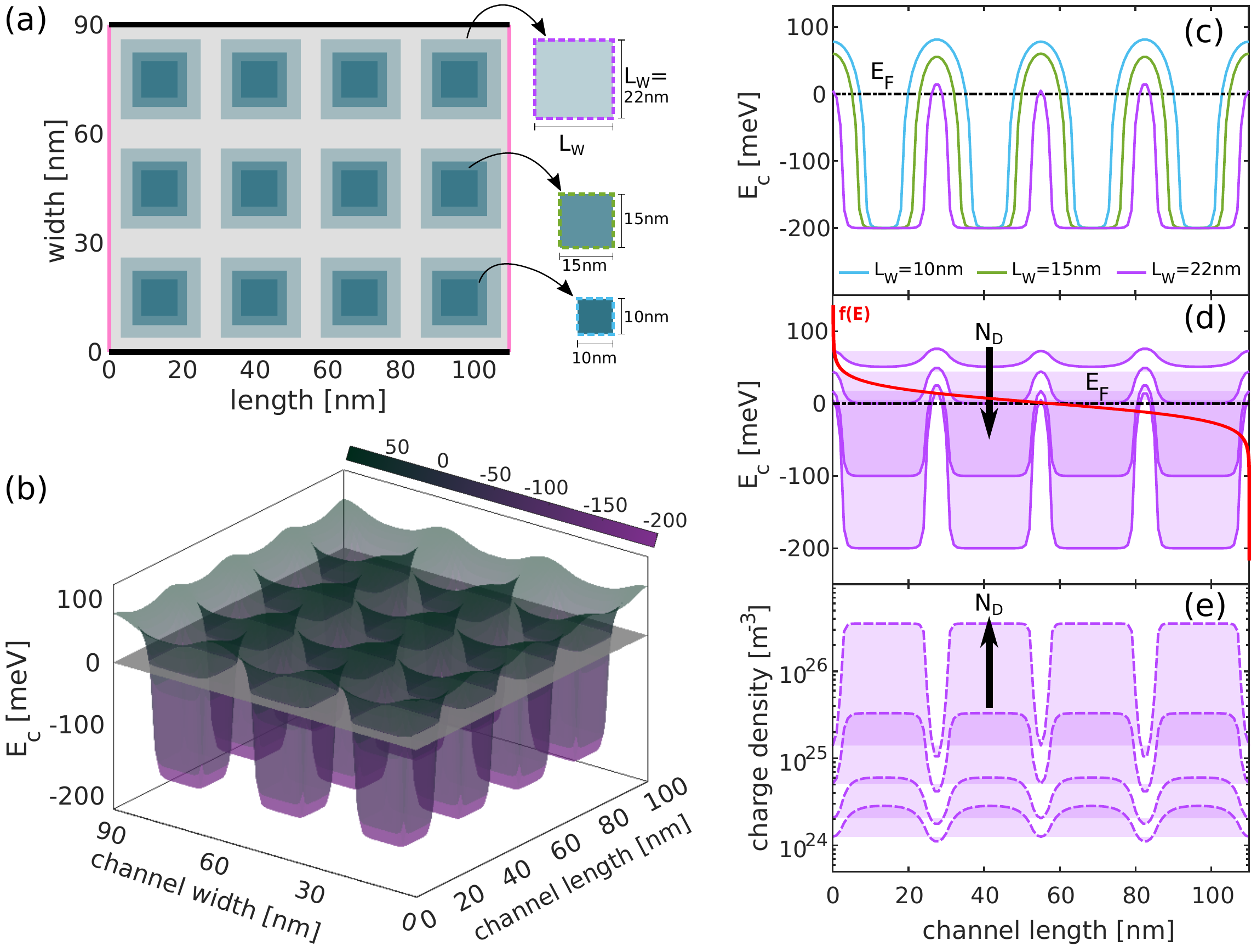}
		\caption{(\textbf{a}) The 2D nanostructure with periodically doped regions (square regions) and dopant-depleted regions simulated. Three sizes for the well regions are considered: $L_W = 10$, $15$, and~$22$ nm. (\textbf{b}) Conduction band $E_c$ profile for $L_W = 22$ nm as calculated from the self-consistent solution of the Poisson equation. (\textbf{c}) A 1D cross-sectional view of the $E_c$ profile along the domain at $45$ nm. (\textbf{d},\textbf{e}) The behavior of the $E_c$ and carrier density for the $L_W = 22$ nm case channel with increasing doping density $N_D$.}
		\label{fig3}
	\end{figure*}

	\indent In a 2D domain featuring periodically arranged doped square sections {(}
	see Figure~\ref{fig1}b{)}, the~spatial profile of $E_c$ across the domain is obtained from the solution of Equation~(\ref{eq1}), with~a typical example shown in Figure~\ref{fig3}b. For~the calculation of the electron flux, we inject electrons at all energies from the left side of the domain and all possible directions. Electrons undergo free flight and scattering events, and~we ray-trace their paths until they exit the domain from the right side. We record the time an~electron spent in the domain and express this as its time of flight (ToF). Electrons backscattered to the left are excluded from flux calculations as they do not contribute. The~inverse of this quantity ensembled from many trajectories provides the flux~\cite{NeophytouBook2020, lundstrom}. We compute the average ToF per particle and utilize it to determine the flux per simulated electron at each energy:
	\begin{equation}
		F(E) = \frac{1}{<ToF(E)>}.
	\end{equation}
	\indent Multiplying the flux per electron by the density of states, $g(E)$, allows us to estimate the transport function as a function of energy.
	\begin{equation}
		\Xi(E) = C \times F(E) \times g(E),
	\end{equation}
	where the~constant $C$ in the TDF equation accounts for two things: the super-electron charge used in Monte Carlo simulations, because~we only simulate a limited number of electrons, and~the geometric factors due to simulating in a finite 2D domain instead of an infinite 3D domain. This constant is determined to map the conductivity computed from Monte Carlo to the bulk material conductivity that we can compute analytically (see~\cite{Pankaj2023} for details). 
	
	This relationship is analogous to the transport distribution function (TDF) $\Xi ( E )$ in the BTE. Essentially, the~product of flux with the density of states represents the flow of charge, which is directly correlated with the conductivity, much like how the TDF influences conductivity~\cite{Pankaj2023}. We calculate the conductivity and the Seebeck coefficient as
	\begin{equation}
		\sigma = q^2 \int \Xi(E) \left( - \frac{\partial f}{\partial E} \right) dE,
		\label{eq3}
	\end{equation}
	\begin{equation}
		S = \frac{q k_B}{\sigma} \int \Xi(E) \left( - \frac{\partial f}{\partial E} \right) \left( \frac{E - E_F}{k_B T} \right) dE,
		\label{eq4}
	\end{equation}
	where $q$ is the electron charge and $k_B$ is the Boltzmann constant, while we set $T = 300$ K in all~simulations.
	
	\section{Power Factor in a Structure with Doped/Intrinsic~Regions}\label{sec3}
	
	We begin our investigation by simulating the TE properties of a 2D nanostructure with dimensions $L = 110~\mathrm{nm}$ (length) and $W = 90~\mathrm{nm}$ (width), with~periodically doped and undoped regions, as~depicted in Figure~\ref{fig3}a. The~Fermi energy is set to $E_F = 0$, and~we consider three sizes of the doped regions: $L_W = 10$, $15$, and~$22$ $\mathrm{nm}$, corresponding to barrier widths of  $L_B \approx 19$, $13$, and~$6$ $\mathrm{nm}$, respectively. We begin by solving the Poisson equation self-consistently for a doping density of $N_D = 3.5 \times 10^{26}$ m$^{-3}$ to determine the profile of the conduction band energy ($E_c$), as shown in Figure~\ref{fig3}b for $L_W = 22$ $\mathrm{nm}$. We deliberately select a high density in accordance with previous studies~\cite{NN_Nanotech_2013,NN_MatToday_2019,Narducci_APL_2021}. In~Figure~\ref{fig3}c, we present a cross-sectional view of the 2D $E_c$ profile at $45~\mathrm{nm}$ across the domain for all three sizes of the doped regions, i.e.,~passing through the middle of the doped wells. Increasing the well size shrinks the barrier region and reduces the barrier height (compare the purple and blue lines in Figure~\ref{fig3}c). Increasing the doping density leads to deeper wells (and lower barriers) and~shifts the top of the barriers toward the Fermi level; see Figure~\ref{fig3}d for the band profile (and Figure~\ref{fig3}e for the carrier density in the $L_W = 22~\mathrm{nm}$ scenario). Thus, the barrier height and~its distance from the Fermi level, which determines the Seebeck coefficient and energy filtering, can be controlled by the geometrical features in the domain (doped/intrinsic region sizes) and the doping~level. 
	
\begin{figure*}
		\includegraphics[height=4cm, width=16cm ]{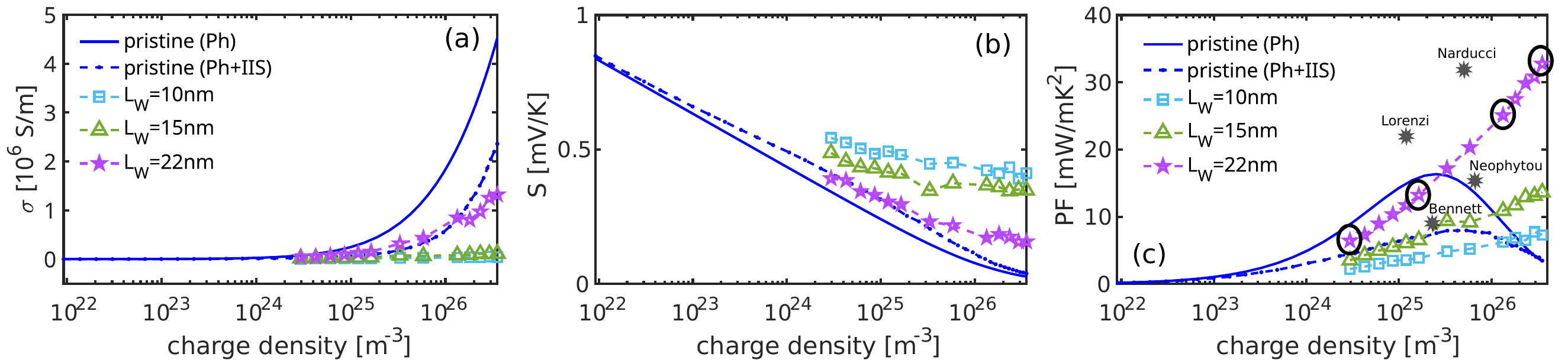}
		\caption{(\textbf{a}) Conductivity, (\textbf{b}) Seebeck coefficient, and~(\textbf{c}) $PF$ versus charge density for (i) pristine material with e-ph (solid blue line) and e-ph plus ionized impurity scattering (blue dashed line) and~(ii) for the periodically doped nanostructures of three different well sizes as indicated. Notably, for~a relatively large well size ($L_W = 22$ nm), the~$PF$ shows a significant increase at high densities. In~(\textbf{c}), the~data points encircled correspond to the charge and potential profiles depicted in Figure~\ref{fig3}d--e.  The gray star marked data show recent experimental measurements from~\cite{NN_Nanotech_2013, Narducci_APL_2021, Bennett_2016, Lorenzi2014}, which show similar power factor enhancements in nanocrystalline materials.}
		\label{fig4}
	\end{figure*}

	\indent For comparison, we first calculate the TE coefficients of the pristine material for the two cases: (i) pristine material subjected only to electron--phonon (e-ph) scattering (representing the ultimate performance of the material) and~(ii) doped pristine material undergoing e-ph and IIS scattering, reflecting the realistically achievable performance of the material. It is worth noting that, for IIS, both the scattering rate and the screening length depend on the doping concentration~\cite{lundstrom}.
	In Figure~\ref{fig4}, we present the conductivities, Seebeck coefficients, and~$PF$s as functions of charge density for these cases. For~the pristine material in the absence of IIS (solid blue lines), the~$PF$ exhibits a peak at high densities (around $10^{25} - 10^{26}$/m$^3$
	), exceeding $2 \times$ the $PF$ in the presence of IIS (dashed blue lines). This is mainly due to the suppression of conductivity, which is primarily due to the stronger influence of IIS compared to phonon-limited scattering, despite the slight increase in the Seebeck~coefficient. 
	
	\indent When the nanostructure is periodically doped, electrons traverse through alternating potential wells and potential barriers; see Figure~\ref{fig3}c. In~this context, we simulate three systems with different well lengths, as~indicated in the legend of Figure~3a. For~each system, we vary the doping density within the range $N_D = 3 \times 10^{24} - 3 \times 10^{26}$/m$^3$. 
	We compare the $PF$ results of these structures with the realistically achieved performance of the pristine material (dashed blue line), as,~in practice, doping facilitates high carrier densities despite reduced mobility due to IIS. We also compare the results of the filtering structures with the solid blue line, the~ultimate performance achievable by our fictitious material. This comparison covers scenarios where gating or modulation doping enables high carrier densities in the absence of dopants in the~lattice.
	
	\indent In the filtering systems, electrons in the wells are not filtered and traverse over the barriers and possess higher energies, indicating a larger Seebeck coefficient. However, the~formation of barriers simultaneously decreases conductivity. This is observed in Figure~\ref{fig4} for $L_W = 10$ and $15~\mathrm{nm}$. Despite the lower conductivity compared to the pristine case, the~corresponding Seebeck coefficient is large, leading to a slight increase in the $PF$ at high densities (around an order of magnitude higher than the densities at which the $PF$ peaks in the pristine case). As~the well size increases to $L_W = 22~\mathrm{nm}$, the~barriers become narrower ($L_B \approx 6~\mathrm{nm}$), and~their height decreases, which mitigates the resistance they introduce. At~such higher densities (as shown in Figure~\ref{fig3}c--e), the~resistance is further mitigated by the use of higher velocity carriers coming from the wells, resulting in a significant increase in conductivity. Note that the Seebeck coefficient decreases as the barrier height reduces, but it still remains considerably higher compared to the pristine material case. This increase in the Seebeck coefficient, in~combination with the relative robustness of the electrical conductivity, leads to high $PF$ improvements. In~fact, the~$PF$ reaches approximately $33$ mW m$^{-1}$K$^{-2}$ for $L_W = 22~\mathrm{nm}$ at a charge density of $7 \times 10^{26}$ m$^{-3}$. In~Figure~\ref{fig4}c, we indicate with black circle the four cases for which we simulate the band profile and carrier density in Figure~\ref{fig3}d--e. Across all cases, the $PF$ is enhanced compared to the pristine material, with~the most significant benefits observed at higher well doping and narrower/reduced barrier height cases. It is important to note that our simulations reveal a monotonic increase in $PF$. This increase will persist as long as the top of the barrier remains above $E_F$. Our final simulation point for high density lies at the borderline, as~shown in Figure~\ref{fig3}d. Beyond~this point, we anticipate a reduction in $PF$, although~we encounter convergence issues in our simulations for such ultra-high doping densities. We also indicate some findings of the PF in Figure~\ref{fig4}c, marked with black stars. These findings align with recent experimental results showing similar power factor enhancements in nanocrystalline materials~\cite{NN_Nanotech_2013, Narducci_APL_2021, Bennett_2016, Lorenzi2014}.
	
	\indent Typically, the~introduction of potential barriers in material systems reduces electrical conductivity, as~carriers encounter exponentially increased difficulty in traversing over the barriers. However, in~the case where $L_W = 22~\mathrm{nm}$ and for the highest density considered (right-most purple line in Figure~\ref{fig4}), the~electrical conductivity is relatively high, reaching up to half of what the pristine material can offer (Figure \ref{fig4}a, comparing the purple and blue-dashed lines at the highest density region). There are a couple of reasons why the conductivity is not substantially degraded despite the barrier regions posing higher resistance: (i) the barrier height, in~that case, is relatively small (as seen with the purple line in Figure~\ref{fig3}c), and~(ii) the barriers are undoped, signifying that carriers entering those regions having high mobility, compensating for the lower velocities of electrons near the band~edge.
	
\begin{figure}
		\includegraphics[height=5cm, width=7cm ]{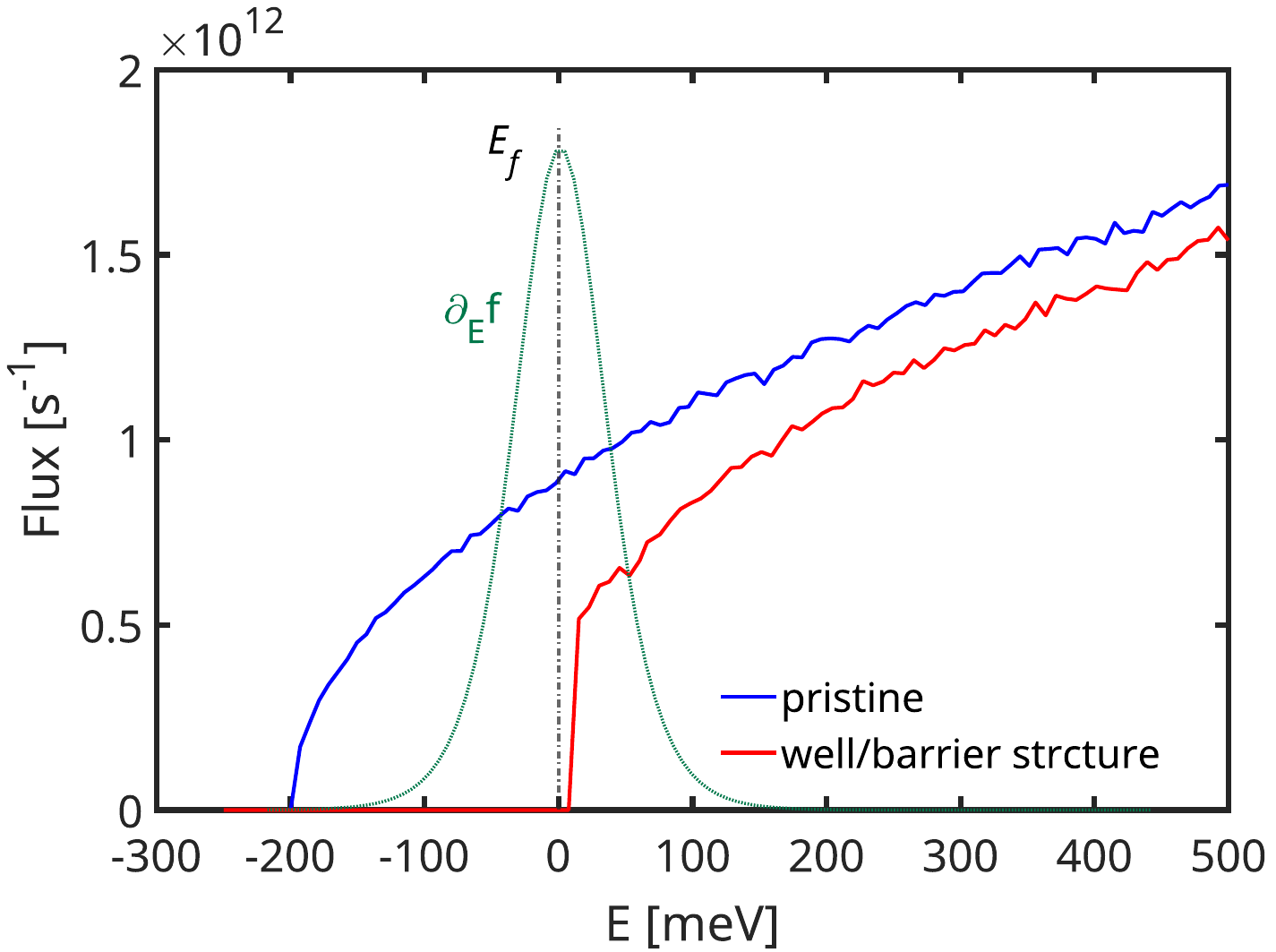}
		\caption{Electron flux extracted from Monte Carlo simulations for the pristine material (blue line) and the nanoengineered channel with periodically doped regions with $L_W = 22$ nm (red line). Additionally, the~derivative of the Fermi distribution is depicted in green, while the Fermi level is represented by the black dashed~line.}
		\label{fig5}
	\end{figure}

	\indent To provide more insight, we illustrate the average flux of a single electron as a function of energy, computed via Monte Carlo, in~Figure~\ref{fig5}. This quantity represents the inverse of the average time of flight taken for an electron to travel from the left to the right contacts of the material system. It considers the different scattering mfp\textit{s} and carrier bandstructure velocities across different regions of the material. We take as reference energy the position of the band profile in the potential well, i.e.,~$E_{C,min} = -200$ meV, and~show two cases. The~flux for a pristine highly doped material (blue line in Figure~\ref{fig5}) and~the flux for the well--barrier system yield the highest reported $PF$ (right-most point on the purple line in Figure~\ref{fig4}c). The~flux for the well--barrier structure (red line in Figure~\ref{fig5}) starts at the energy of the barrier height, which is a few meV above $E_F = 0$ eV in this case. A~key observation is a sharp increase in flux after encountering the barrier, reaching approximately 70 percent of the value for the pristine material within a few meV. This observation explains the limited degradation of conductivity by potential barriers in this specific case, as~observed in Figure~\ref{fig4}a.
	
	\section{Summary and~Conclusions}\label{sec4}
	
	In this study, we investigated the thermoelectric properties of nanoengineered materials featuring highly doped regions periodically separated from undoped regions. Employing Monte Carlo simulations for electronic transport, coupled self-consistently with the Poisson equation to capture the electrostatics of the domain, we demonstrated the potential of these structured materials to achieve thermoelectric power factors up to five times higher than the optimal power factor of pristine materials. These findings, derived from advanced numerical simulations and software, confirm predictions made by much simpler models~\cite{NN_MatToday_2019} and~are in line with recent experimental results showcasing such power factor enhancements in nanocrystalline materials~\cite{NN_Nanotech_2013,Narducci_APL_2021,Lorenzi2014}. Our results can motivate the significance of further research into nanostructured thermoelectric materials aiming to achieve ultra-high power factors through efficient utilization of energy filtering~mechanisms. 

		
	\bibliographystyle{apsrev}
	\bibliography{Reference}	
\end{document}